\documentstyle[a4,11pt,epsfig,twoside]{article}
\begin{document}
\setcounter{page}{1}
\title{Studying the Higgs Sector\\ at the 
CLIC Multi-TeV $e^+e^-$ Collider}

\author{M. Battaglia
        \thanks{e-mail address: Marco.Battaglia@cern.ch}
        and A.~De~Roeck
        \thanks{e-mail address: Albert.De.Roeck@cern.ch}
\\
\\
        {\it CERN, Geneva, Switzerland}
}
\date{}
\maketitle
\begin{abstract}
In this paper, we review the role of a multi-TeV $e^+e^-$ linear collider to complete 
the mapping of the Higgs boson profile and studying heavier bosons in extended scenarios
with more than one Higgs doublet.
\end{abstract}

\section{Introduction}

Understanding the origin of electro-weak symmetry breaking and mass generation stands 
as a central theme of the research programme in physics in the coming decades.
The Standard Model (SM), successfully tested to an unprecedented level of accuracy 
by the LEP and SLC experiments, and now also by the $B$-factories, addresses 
this question with the Higgs mechanism~\cite{Higgs}. 
The first manifestation of the Higgs mechanism through the Higgs sector is the 
existence of at least one Higgs boson, $H^0$. The observation of a new 
spin-0 particle would represent a first sign that the Higgs mechanism of mass 
generation is realised in Nature. Present data indicates that it is 
heavier than 114~GeV and possibly lighter than about 195~GeV~\cite{ichep}. 
We expect the Higgs boson to be discovered at the {\sc Tevatron} or at the 
{\sc Lhc}, the CERN hadron 
collider, which will determine its mass and perform a first survey of its basic 
properties. A TeV-class linear collider, operating at centre-of-mass energies 
350~GeV$< \sqrt{s} \le$~1~TeV, will bring the accuracy needed to further validate 
this picture and to probe the SM or extended nature of the Higgs sector. It 
will perform crucial measurements in a model-independent way and it will also complement 
the {\sc Lhc} in the search for heavy Higgs bosons. Their observation would 
provide direct evidence that nature has chosen a route different than the minimal 
Higgs sector of the SM. 
But neither the precision study of the Higgs profile nor the search for
additional Higgs bosons will be completed at energies below 1~TeV. There are 
measurements which will be limited in accuracy or may be not feasible at all, due to 
limitations in both the available statistics and centre-of-mass energy. The two-beam 
acceleration scheme, presently developed within the {\sc Clic} study~\cite{clic} 
at CERN, aims at collisions at $\sqrt{s}$=3-5~TeV, representing a unique opportunity to 
extend $e^+e^-$ physics to constituent energies of the order of the {\sc Lhc} energy 
frontier, and beyond, with very high luminosity. It is therefore important to assess the
potential of a multi-TeV LC in complementing information that the 
{\sc Lhc}, and possibly a lower energy LC, will obtain.

\section{Completing the Light Higgs Boson Profile}

The TeV-class LC will perform highly accurate determinations of the Higgs 
profile. However, even at the high design luminosity of {\sc Tesla} and the {\sc Nlc} 
there are properties which cannot be tested exhaustively. 
\begin{figure}[h!]
\vspace*{-0.75cm}
\begin{center}
\epsfig{file=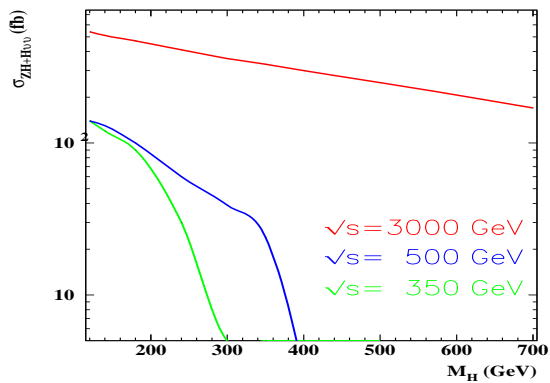,width=7.5cm,height=5.5cm,clip}
\end{center} 
\vspace*{-0.75cm}
\caption[]{\sl Inclusive Higgs production cross section as a function of the Higgs mass 
$M_H$ for three values of the $e^+e^-$ centre-of-mass energy $\sqrt{s}$.} 
\label{fig:xsec}
\end{figure}
In fact, it is essential to ensure that the fundamental test of the scaling of Higgs 
couplings to fermions with their masses 
$\frac{g_{HXX}}{g_{HYY}} \propto \frac{M_X}{M_Y}$ can be performed with significant 
accuracy over a wide range of Higgs boson masses and for all particle species. 
At $\sqrt{s}=350-500$~GeV, the LC will test the couplings to gauge boson, and those to 
quarks if the Higgs boson is light. To complete this program for leptons and 
intermediate masses of the Higgs and to study the Higgs boson self-couplings, it is 
necessary to study rare processes, which need Higgs samples in excess to $10^5$ events.
As the cross section for $e^+e^- \rightarrow H^0 \nu \bar \nu$ production increases with 
energy as $log~\frac{s}{M^2_H}$, it dominates at {\sc Clic} energies. 
The resulting large Higgs production rate at $\sqrt{s} \ge$~3~TeV and the expected 
luminosity, $L = 10^{35}$~cm$^{-2}$s$^{-1}$, yield samples of order of 
0.5-1~$\times 10^6$ decays of SM-like Higgs boson in 1-2 years 
(see Figure~\ref{fig:xsec}).

Measuring the muon Yukawa coupling by the determination of the 
$H^0 \rightarrow \mu^+\mu^-$ branching fraction completes the test of the coupling 
scaling for gauge bosons, quarks and leptons separately. This ensures that the observed 
Higgs boson is indeed responsible for the mass generation of all elementary particles.
At $\sqrt{s}$=3~TeV, the $H \rightarrow \mu^+ \mu^-$ signal can be easily observed and 
the $g_{H\mu\mu}$ coupling measured to 4-11\% accuracy for Higgs masses in the range 
120~GeV~$< M_H <$~150~GeV, with 3~ab$^{-1}$. 
\begin{figure}
\vspace*{-0.75cm}
\begin{center}
\begin{tabular}{c c}
\epsfig{file=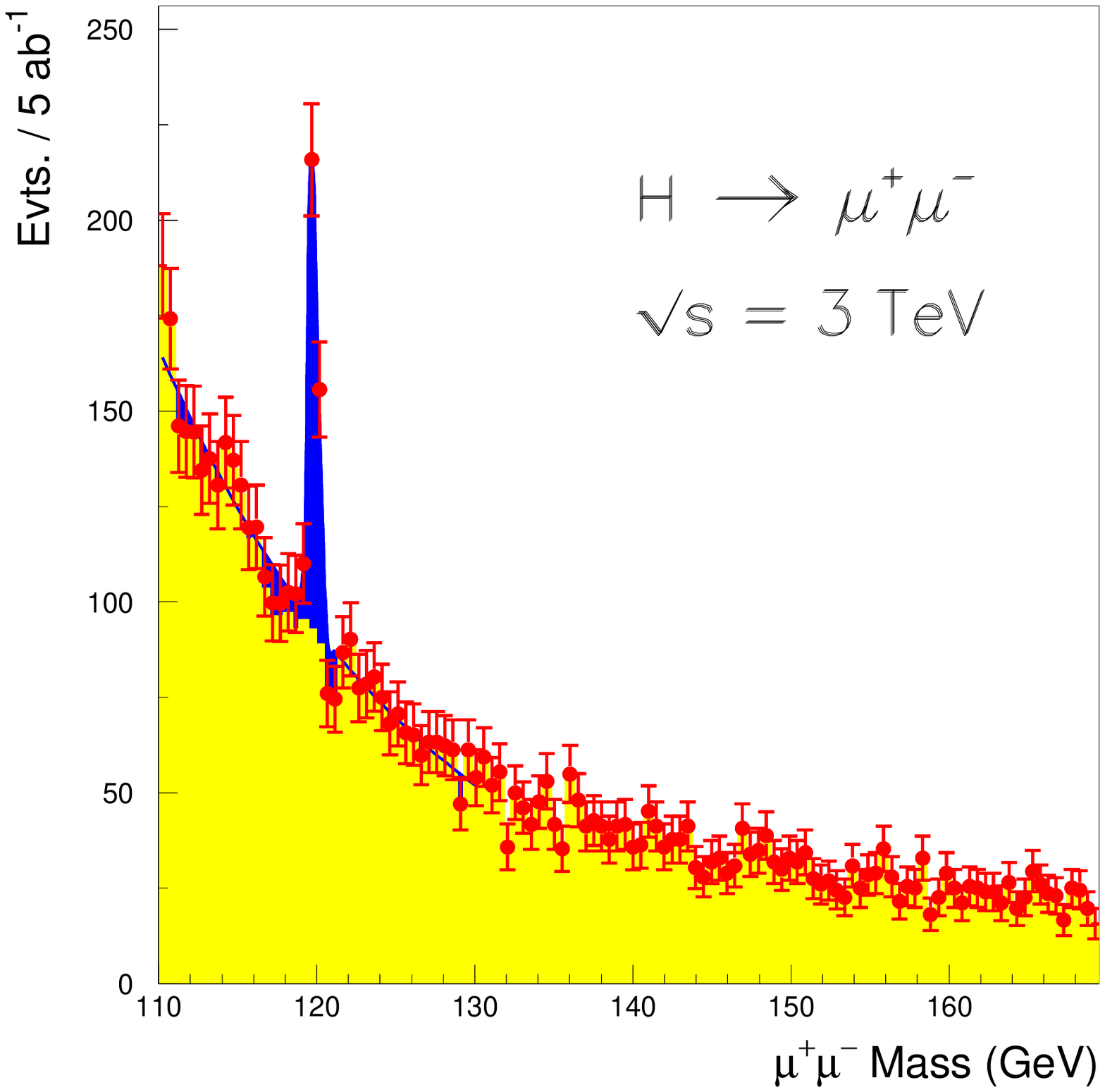,width=6.75cm,height=5.0cm,clip} &
\epsfig{file=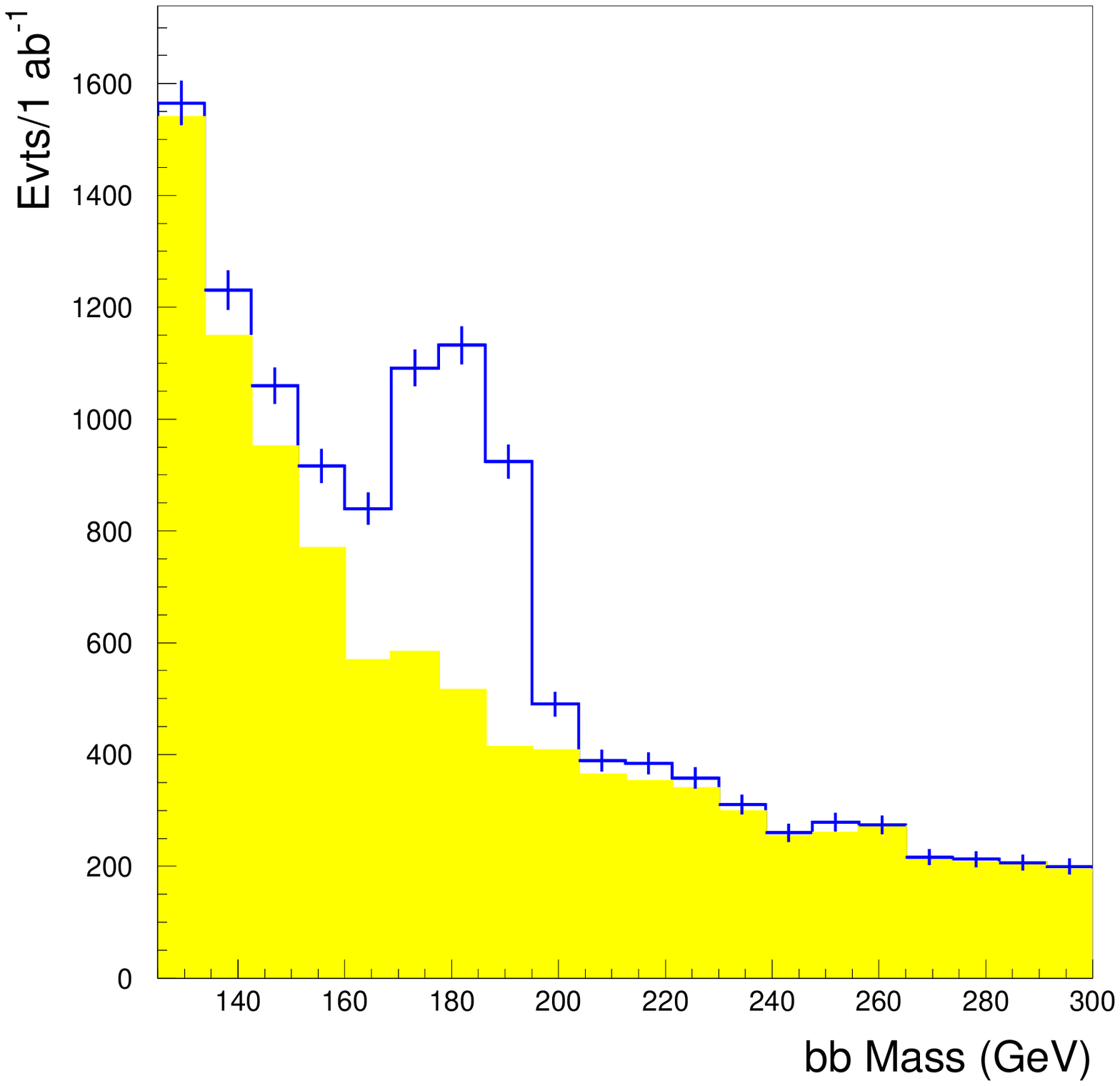,width=6.75cm,height=5.0cm} \\
\end{tabular}
\end{center}
\vspace*{-0.75cm}
\caption[]{\sl Reconstructed signals for $e^+e^- \to H^0 \to  \mu^+\mu^-$ (left) and
 $e^+e^- \to H^0 \to  b\bar{b}$ (right) for $M_H$ = 120~GeV and 200~GeV respectively at
$\sqrt{s}$=3~TeV.}
\label{fig:rare}
\end{figure}
\begin{table}
\caption[]{\sl Signal significance and relative accuracy for the determination of the 
$b$ Yukawa coupling for different values of the $M_H$ mass at $\sqrt{s}$=3~TeV.}
\begin{center}
\begin{tabular}{|c|c|c|c|}
\hline
$M_H$ (GeV) & $S/\sqrt{B}$ & $\delta~g_{Hbb}~/~g_{Hbb}$ \\
\hline \hline
180 & 40.5 & 0.016 \\
200 & 25.0 & 0.025 \\
220 & 18.0 & 0.034 \\ \hline
\end{tabular}
\end{center}
\vspace*{-0.50cm}
\label{tab:bb}
\end{table}
The sensitivity on the Higgs couplings to fermions also needs to be tested for heavier 
Higgs bosons. Beyond the $H \to WW$ threshold, the branching 
fractions $H \to f \bar{f}$ fall rapidly with increasing $M_H$ values. 
$e^+e^- \rightarrow \nu \bar \nu H \rightarrow b \bar b$ at $\sqrt{s} \ge$ 
1~TeV offers a favourable signal-to-background ratio to probe $g_{Hbb}$ for these 
intermediate-mass Higgs bosons. The $H^0 \to b \bar{b}$ decay can be measured for 
masses up to about 240~GeV and the accuracies are summarised in Table~\ref{tab:bb}.

Another fundamental test of the Higgs sector with a light Higgs boson, which 
significantly benefits from multi-TeV data, is the study of the Higgs self-couplings and 
the reconstruction of the Higgs potential.
\begin{table}
\caption[]{\sl Relative accuracy for the determination of the 
triple Higgs coupling $g_{HHH}$ for different values of the $M_H$ mass at 
$\sqrt{s}$=3~TeV, assuming unpolarised beams.}
\begin{center}
\begin{tabular}{|c|c|c|}
\hline
$M_H$ (GeV) & Counting & Fit \\ \hline
120 & $\pm$ 0.131 (stat) & $\pm$ 0.093 (stat) \\
180 & $\pm$ 0.191 (stat) & $\pm$ 0.115 (stat) \\ \hline
\end{tabular}
\end{center}
\vspace*{-0.50cm}
\label{tab:hhh}
\end{table}
The triple Higgs coupling $g_{HHH}$ can be accessed at a TeV-class LC in the double 
Higgs production processes $e^+e^- \to H H Z$~\cite{Djouadi:1999gv}. This measurement 
is made difficult by the tiny production cross section and by the 
dilution due to diagrams leading to double Higgs production, but not sensitive to the 
triple Higgs vertex. A LC operating at $\sqrt{s}$ = 500~GeV can measure 
the $HHZ$ production cross section to about 15\% accuracy if the Higgs boson mass is 
120~GeV, corresponding to a fractional accuracy of 23\% on 
$g_{HHH}$~\cite{Castanier:2001sf}. Improvements can be obtained by performing the 
analysis at multi-TeV energies, through the process $e^+e^- \to H H \nu \bar{\nu}$ and 
by introducing observables sensitive to the presence of the triple Higgs vertex  
(see~Table~\ref{tab:hhh})~\cite{Battaglia:2001nn}. 
On the contrary, the quartic Higgs coupling remains elusive, due to the smallness of 
the relevant triple Higgs production cross sections.

Precision electro-weak data indicate that the Higgs boson must be lighter than about 
195~GeV. However, this limit can be evaded if New Physics exists to cancel the effect 
of the heavy Higgs boson mass. In these scenarios, it is interesting to search for an
heavier boson through the $ZZ$ fusion process $e^+e^- \to H^0 e^+ e^- \to X e^+e^-$, 
at high energies. Similarly to the associate $HZ$ production in 
the Higgs-strahlung process at lower energies, this channel allows a model-independent 
search of Higgs boson, through the tag of the two forward electrons and the 
reconstruction of their recoil mass~\cite{Battaglia:2001yi}. This analysis needs to 
identify electrons, and measure their energy and direction down to $\simeq 100$~mrad, 
close to bulk of the $\gamma\gamma \to {\mathrm{hadrons}}$ and pair backgrounds, making 
it a challenge for the forward tracking and calorimetric response of the detector. 
Preliminary results show that a clean Higgs signal can be extracted at $\sqrt{s}$=3~TeV 
for $M_H \le$~900~GeV.

\section{Testing New Physics in the Higgs Sector}

If heavy Higgs bosons exist above pair-production threshold, they are accessible
at the LC through the $e^+e^- \to H^0$ $A^0 \to b \bar{b} b \bar{b}$, 
$t \bar{t} t \bar{t}$ and $e^+e^- \to H^+H^- \rightarrow t \bar{b} \bar{t} b$
processes, resulting in very distinctive, yet challenging, multi-jet final states with 
multiple $b$-quark jets, which must be efficiently identified and reconstructed. 
\begin{figure}[h!]
\vspace*{-0.75cm}
\begin{center}
\begin{tabular}{c c c}
\epsfig{file=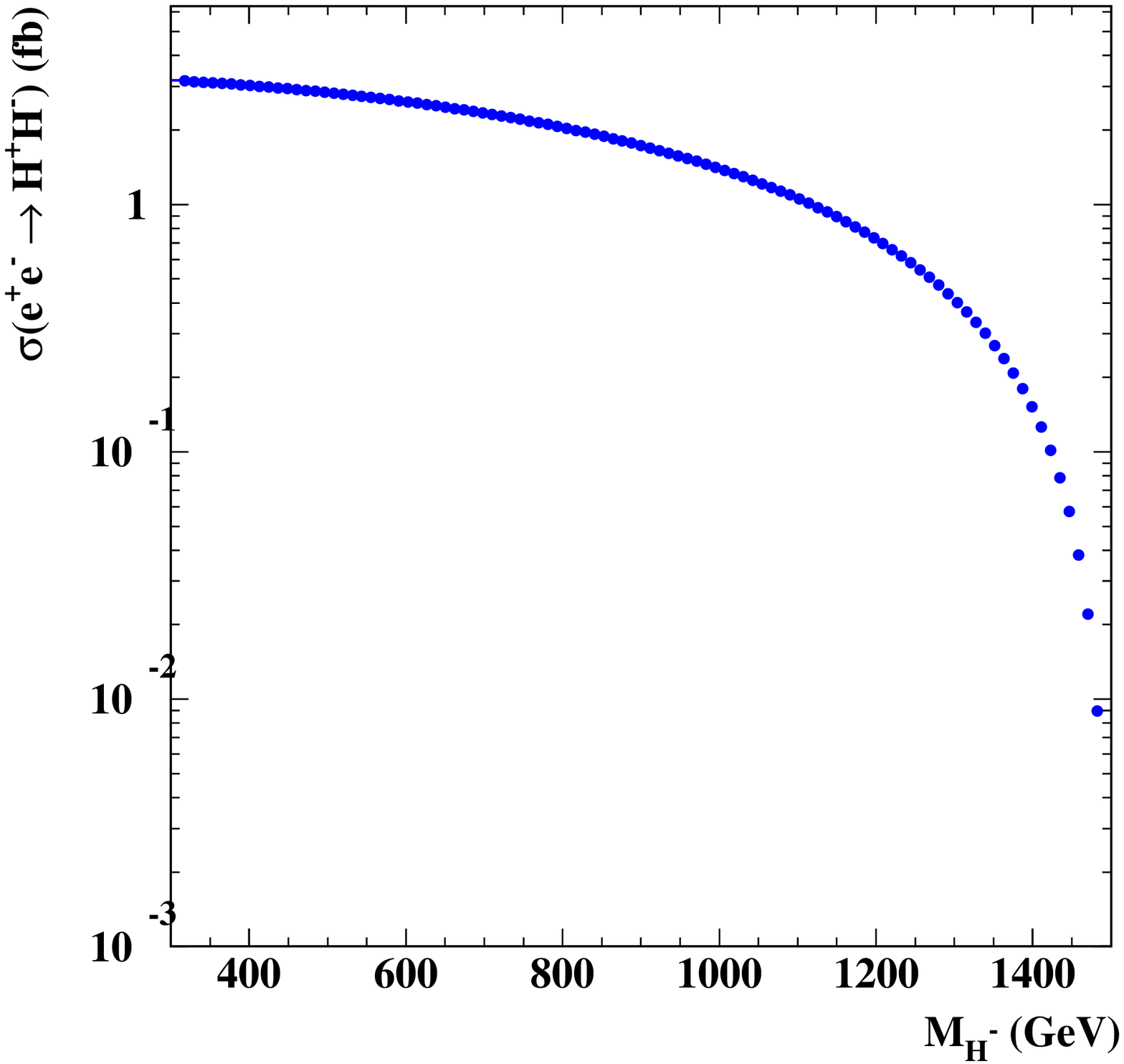,width=4.0cm,height=5.0cm} &
\epsfig{file=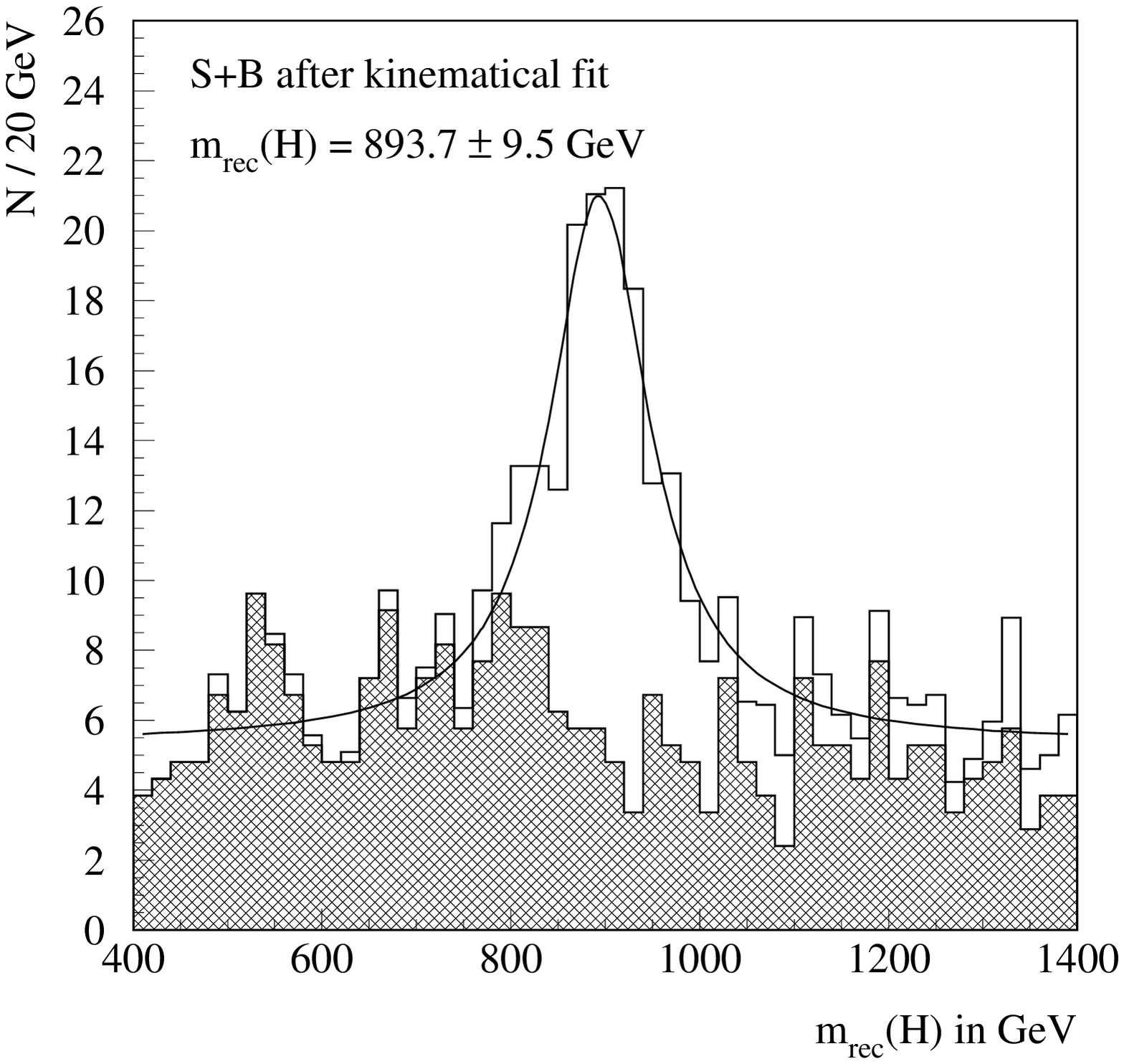,width=4.5cm,height=4.8cm} &
\epsfig{file=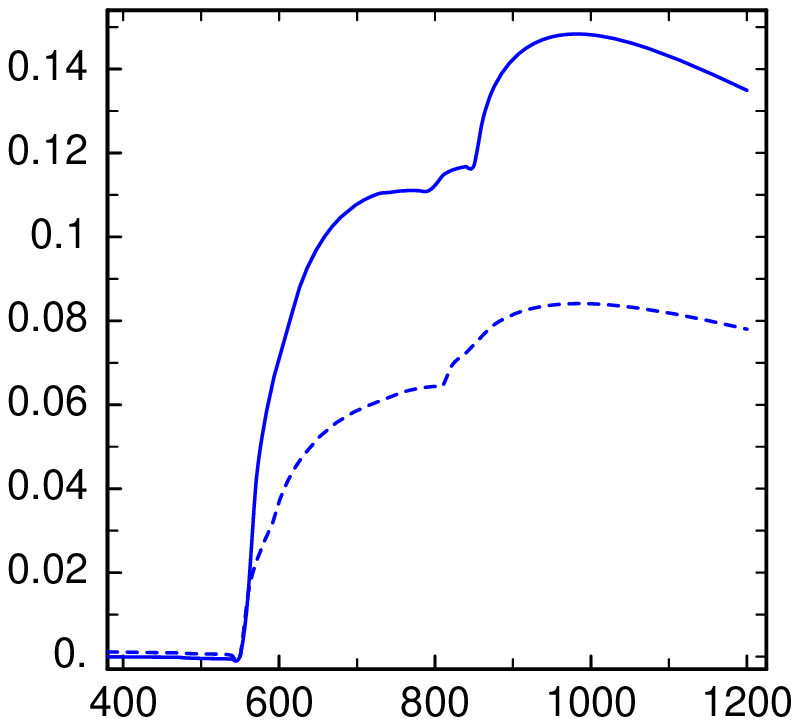,width=4.0cm,height=5.0cm} \\
\end{tabular}
\end{center}
\vspace*{-0.75cm}
\caption[]{\sl Charged Higgs analysis at {\sc Clic}: $e^+e^- \to H^+H^-$ production 
cross section for $\sqrt{s}$=3~TeV as function of the charged Higgs mass (left). Signal 
reconstructed in the $H^+ \to t \bar{b}$ decay mode with the irreducible background 
overlayed (centre). Magnitude of the CP asymmetry $|\delta^{CP}|$ estimated as function 
of the charged Higgs mass (right).}
\label{fig:hpm}
\end{figure}
The sizeable production cross sections provides sensitivity for masses up to 1~TeV and 
beyond for all values of $\tan \beta$, thus extending the LHC reach. A detailed analysis
has been performed for the reconstruction of $e^+e^- \to H^+H^- \to t \bar{b} \bar{t} b$
with $M_H$=880~GeV, corresponding to the CMSSM benchmark point~J of 
reference~\cite{cmssm}, by identification of the 
$W bb W bb$ final state, applying a mass constrained fit~\cite{hpm}. The irreducible 
SM $e^+ e^- \to t \bar{b} \bar{t} b$ background and the overlay of accelerator-induced
$\gamma \gamma \to {\mathrm{hadrons}}$ events have been included.
By reconstructing either both $H^{\pm}$s, or only one, the analysis can be optimised in 
terms of efficiency and resolution for mass measurement, or for an unbiased study of 
$H^{\pm}$ decays, respectively. Fractional accuracies of 1\% on the heavy boson mass and 
of 5-10\% on the product of production cross section and $tb$ decay branching fraction 
are expected, with ${\cal{L}}$=3~ab$^{-1}$ at $\sqrt{s}$ = 3~TeV.

Extensions of the SM may introduce new sources of CP violation, through additional 
physical phases whose effects can be searched for in the Higgs sector.
Supersymmetric one-loop contributions can lead to differences in the decay rates of 
$H^+ \to t \bar{b}$ and $H^- \to \bar{t} b$, in the MSSM with complex 
parameters~\cite{Christova:2002ke}. This CP asymmetry is expressed as: 
$|\delta^{CP}| = 
\frac{|\Gamma(H^- \rightarrow b\bar{t})-\Gamma(H^+ \rightarrow t\bar{b})|}
{\Gamma(H^- \rightarrow b\bar{t})+\Gamma(H^+ \rightarrow t\bar{b})}$ and it can amount 
to up to $\simeq 15$\% as shown in Figure~\ref{fig:hpm}. 
The leading contributions, from loops with $\tilde{t}$, $\tilde{b}$ and 
$\tilde{g}$, $\delta^{CP}$, are sensitive to the masses of these sparticles.
With the expected statistics of $e^+e^- \to H^+H^- \to t \bar{b} \bar{t} b$ at 
$\sqrt{s}$=3~TeV and assuming realistic charge tagging performances, a 3~$\sigma$ effect
would be observed with ${\cal{L}}$=5~ab$^{-1}$, for an asymmetry $|\delta^{CP}|$=0.10.

\begin{figure}[h!]
\vspace*{-0.25cm}
\begin{center}
\begin{tabular}{c c}
\epsfig{file=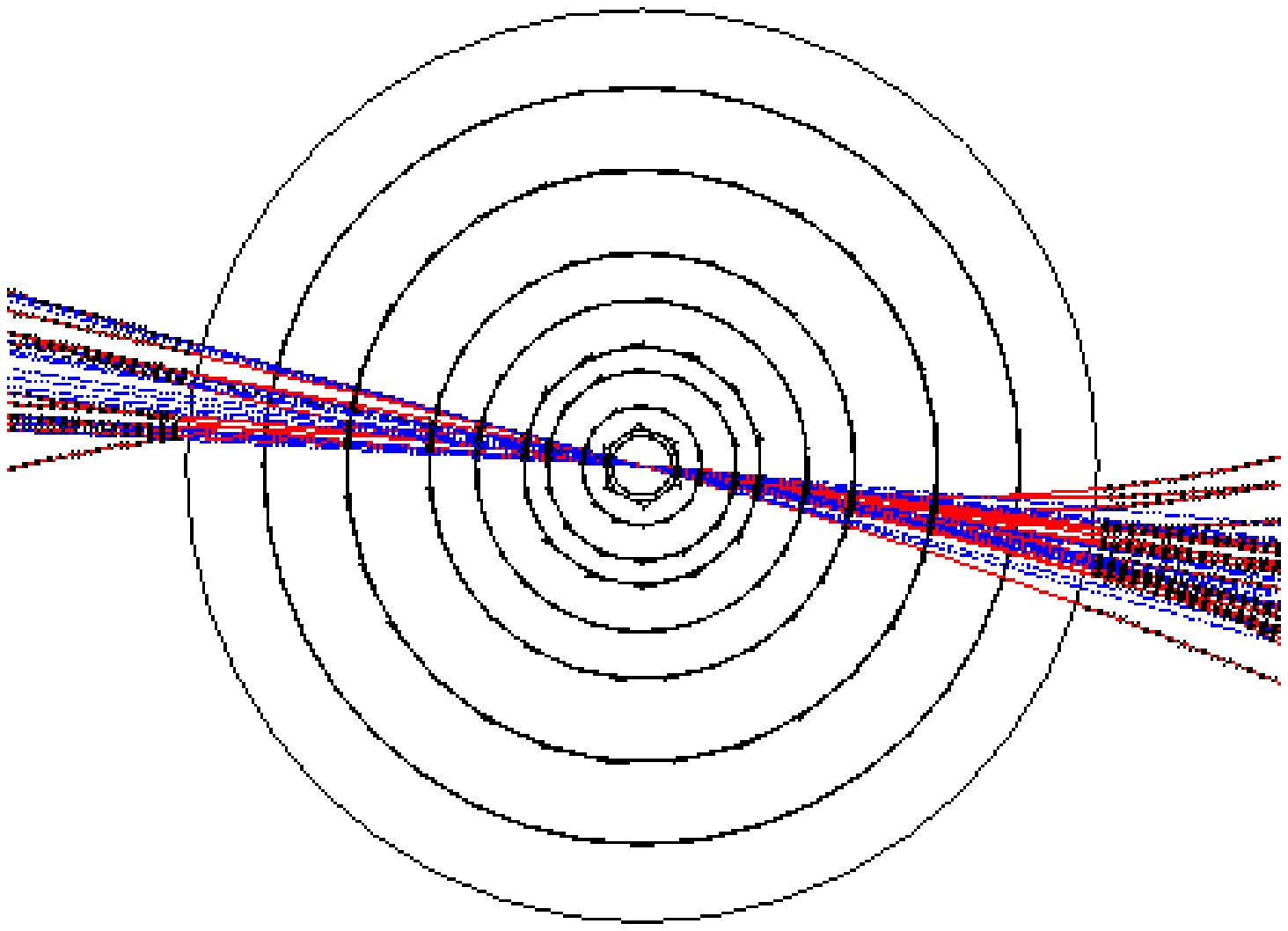,width=6.5cm} &
\epsfig{file=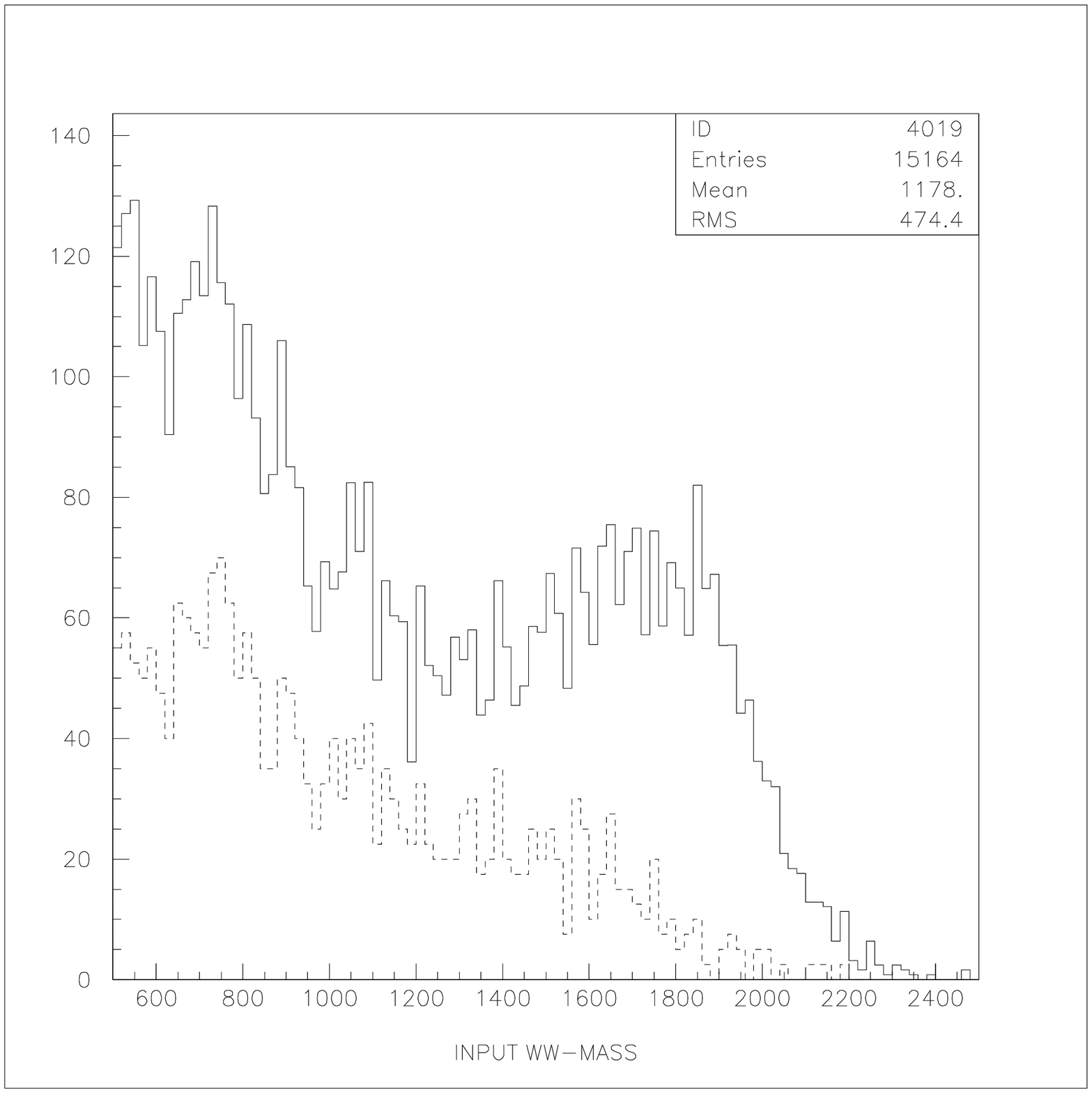,width=4.75cm} \\
\end{tabular}
\end{center}
\vspace*{-0.5cm}
\caption[]{\sl Left: Display of a $e^+e^- \to W^+W^- \nu \bar{\nu}$ event at 3~TeV 
reconstructed in a multi-layered Si Tracker. Right: Distribution of the reconstructed 
$WW$ mass in a SSB model with a 2~TeV resonance. The background is overlayed.}
\label{fig:ww}
\vspace*{-0.5cm}
\end{figure}

\section{EWSB without the Higgs Boson}

If the breaking of the electro-weak symmetry is not due to the Higgs mechanism, then 
the $e^+e^- \rightarrow W^+W^-\nu\bar{\nu}$ and $Z^0Z^0\nu\bar{\nu}$ 
processes may reveal new dynamics of gauge boson interactions~\cite{Barklow:2001is}.
In fact when no elementary Higgs boson exists with $M_H \le$ 700~GeV, we expect 
$W^{\pm}$ and $Z^0$ bosons to develop strong interactions at a scale $\simeq$ 1~TeV.
The experimental signatures are represented both by deviations of the 
$e^+e^- \to W_LW_L\nu\nu$ cross section from its SM expectation and also by the 
possible formation of vector resonances at masses beyond 1~TeV 
(see Figure~\ref{fig:ww}). 
Clean final states can be reconstructed using $W \rightarrow q \bar q'$ and the
resonant components identified also when accounting for backgrounds and detector 
response~\cite{Barklow:2001mm}.

\section{Conclusions}

A Multi-TeV $e^+e^-$ LC, such as {\sc Clic}, has the potential to complete the 
study of the Higgs boson and to investigate an extended Higgs sector over a wide 
range of model parameters.
Preserving the LC signature properties of clean events, with well defined 
kinematics, at a Multi-TeV LC will require a substantial effort
of machine parameter optimisation, detector design and data analysis techniques.
However, exploratory studies, accounting for realistic experimental conditions, 
confirm that {\sc Clic} will perform precision measurements and push its sensitivity 
up to the kinematical limits.
\begin{table}[h!]
\caption[]{\sl Comparison of expected accuracies on Higgs couplings at {\sc Clic} and 
at the {\sc VLhc} hadron colliders~\cite{Baur:2002ka}}
\begin{center}
\begin{tabular}{|c|c|c|c|}
\hline
      & CLIC & VLHC-I & VLHC-II \\
$\sqrt{s}$       & 3~TeV & 40~TeV & 200~TeV \\
$\int{\cal{L}}$  & 5 ab$^{-1}$ & 300 fb$^{-1}$ & 300 fb$^{-1}$ \\ \hline
$\delta g_{Htt}/g_{Htt}$ & 0.05-0.10 (?) & 0.05-0.10 & 0.01-0.02 \\
$\delta g_{Hbb}/g_{Hbb}$ & 0.01-0.03 & - & - \\
$\delta g_{H\mu\mu}/g_{H\mu\mu}$ & 0.03-0.10 & 0.15-0.25 & 0.10-0.13 \\
$\delta g_{HHH}/g_{HHH}$   & 0.07-0.09 & ??? & 0.10-0.30 (?) \\
$g_{HHHH}$ & $\ne 0$ (?) & - & - \\
\hline
\end{tabular}
\end{center}
\vspace*{-0.5cm}
\label{tab:summary}
\end{table}
While the main motivation for experimentation at a Multi-TeV LC arise from 
the search of new phenomena, its role in studying the Higgs sector will also be 
crucial in completing the mapping of the $H^0$ Boson profile, studying heavy Higgs 
bosons in extended scenarios or explore otherwise the origin of symmetry breaking, 
if no elementary Higgs boson is observed. 
In this way, the {\sc Clic} multi-TeV LC project ensures a continue competitive 
$e^+e^-$ physics program through two generation of projects and of physics questions.

{\sl We are grateful to K.~Desch, S.~Kraml and A.~Ferrari for contributions.}

\end{document}